\documentclass[aip,reprint]{revtex4-2}
\usepackage{graphicx}
\usepackage{dcolumn}
\usepackage{bm}
\usepackage[utf8]{inputenc}
\usepackage[T1]{fontenc}
\usepackage{mathptmx,amsmath,amssymb}
\usepackage{mathrsfs}
\usepackage{gensymb}
\usepackage{amsfonts}
\usepackage{amsthm}
\usepackage{natbib}
\usepackage{color}
\usepackage{hyperref}
\usepackage{bm}
\usepackage[caption=false]{subfig}
\usepackage{verbatim}
\usepackage{ulem}
\usepackage{babel}

\begin{document}

	\title{Non-reciprocity of Vortex-limited Critical Current in Conventional Superconducting Micro-bridges}
	\author{Dhavala Suri}
	\email{d.suri@tum.de}
	\affiliation{Department of Physics, Technical University of Munich, Garching 85748, Germany} \author{Akashdeep Kamra}
    \affiliation{Condensed Matter Physics Center (IFIMAC) and Departamento de Física Teórica de la Materia Condensada, Universidad Autónoma de Madrid, E-28049 Madrid, Spain}
	\author{Thomas N. G. Meier}
	\affiliation{Department of Physics, Technical University of Munich, Garching b. Munich, Germany} 
	\author{Matthias Kronseder}
	\affiliation{Department of Physics, University of Regensburg, Regensburg, Germany}
	\author{Wolfgang Belzig}
	\affiliation{Fachbereich Physik, Universitat Konstanz, D-78457 Konstanz, Germany}
	\author{Christian H. Back}
	\affiliation{Department of Physics, Technical University of Munich, Garching b. Munich, Germany}
	\affiliation{Munich Center for Quantum Science and Technology (MCQST),~D-80799 M$\ddot{u}$nchen,~Germany}
	\affiliation{Centre for Quantum Engineering (ZQE), Technical University~of~Munich,~Garching,~Germany}
	\author{Christoph Strunk}
	\affiliation{Department of Physics, University of Regensburg, Regensburg, Germany}

\begin{abstract}  
Non-reciprocity in the critical current has been observed in a variety of superconducting systems and has been called the superconducting diode effect. The origin underlying the effect depends on the symmetry breaking mechanisms at play. We investigate superconducting micro bridges of  NbN and also NbN/magnetic insulator (MI) hybrids. We observe a large diode efficiency  of $\approx$~30\% when an out-of-plane magnetic field as small as 25~mT is applied. In both NbN and NbN/MI hybrid, we find that the diode effect vanishes when the magnetic field is parallel to the sample plane. Our observations are consistent with the critical current being determined by the vortex surface barrier. Unequal barriers on the two edges of the superconductor strip result in the diode effect. Furthermore, the rectification is observed up to a temperature $\sim$10~K, which makes the device potential for diode based applications over larger temperature range than before. 
\end{abstract}
	
\maketitle

In the presence of symmetry breaking potentials, the magnitudes of critical currents of a superconductor (SC) are unequal for  the two  bias polarities  \cite{Tokura2018,nagaosa2018,shintaro2018,Ando2020,ideue2020,Baumgartner2022,heng2022,Strambini2022}.   This phenomenon is called the superconducting diode effect (SDE) and may arise due to simultaneous breaking of time reversal symmetry (TRS) and inversion symmetry (IS). The effect has  gained attention in recent times for its potential applications in non-dissipative electronics. Assuming the critical current of SC to be determined by the critical depairing mechanism, theoretical models for the SDE rely on out-of-plane Rashba spin-orbit coupling \cite{noah2022,bergeret2022,daido2022,karabasov2022,He2022}, valley-Zeeman interaction \cite{bauriedl2022} etc.~ for the IS breaking and an applied magnetic field for the TRS breaking. As a result, SDE in SC is related to emergence of a chiral superconducting order~\cite{noah2022,bergeret2022,daido2022,He2022}.

The field witnessed an upsurge of reports spanning a range of systems such as  van der Waals material with noncentrosymmetric crystal potential -- MoS$_2$ \cite{wakatsuki2017}, synthetic super lattice of Nb/V/Ta \cite{Ando2020}, planar Josephson junction arrays of Al on InAs \cite{Baumgartner2022,Baumgartner_2022b}, magnetic proximity coupled hetero-structures of van der Waals materials \cite{shin2021} etc. These studies of the SDE rely  primarily on the combination of magnetic fields and  spin-orbit coupling, giving rise to magnetochiral anisotropy \cite{pal2021,lin2021,Yasuda2019,itahashi2020,Baumgartner2022,Ando2020,rikken2001}. Furthermore, it has been reported  that an out-of-plane magnetic field can cause the diode effect due to valley-Zeeman spin-orbit interaction in NbSe$_2$ \cite{bauriedl2022} or the imbalance in valley occupation in the case of  twisted tri-layer graphene \cite{lin2021}. 
A recent experimental report has demonstrated non-reciprocal critical current in superconductor films as well as their hybrid with magnetic insulator employing small to no magnetic fields \cite{hou2022}. Via careful experiments, spin-orbit coupling was ruled out as the origin of the SDE. Instead, IS breaking  in those experiments is provided by the non-identical edges of the superconducting film. 

While a majority of the theoretical work has focused on the critical depairing mechanism, the critical current in type II superconductors is often determined by the vortex surface barriers~\cite{Shmidt1970,Shmidt1970b,Rocci2021,Hope2021}. The supercurrent tries to pull vortices  nucleated on one edge towards the other side, where they can be annihilated. At low supercurrents, the Bean-Livingston surface barrier \cite{Bean1964} is strong enough to prevent the vortices from entering the superconductor. However, at a large enough current, which becomes the critical current, the Lorentz force overcomes the surface barrier resulting in flux flow through the film and destruction of the superconducting state~\cite{Shmidt1970,Shmidt1970b}. In case of a thin film, the lowest surface barrier is typically offered by the side surfaces for out-of-plane vortices resulting in a critical current significantly lower than the critical depairing current. Considering this mechanism, unequal vortex barriers owing to asymmetric local defects on the two side surfaces have been predicted to result in the SDE \cite{vodolazov2005}.

In this article, we find SDE in NbN micro-bridges in out-of-plane magnetic field. Furthermore, to investigate the effects of an in-plane exchange field, we study NbN/YIG and find no diode effect with in-plane applied fields. Since our device geometry avoids fringe fields caused by YIG from entering NbN, the absence of diode effect provides a test of the fringe-field mechanism of SDE put forth by Hou and coworkers~\cite{hou2022}. We attribute our observations to the critical current being determined by the vortex flow. Our results confirm that the SDE is caused by unequal vortex barriers on the two side surfaces of the film~\cite{vodolazov2005,Hope2021}.

\begin{figure}[tbh]
\centering
\includegraphics[width=\linewidth]{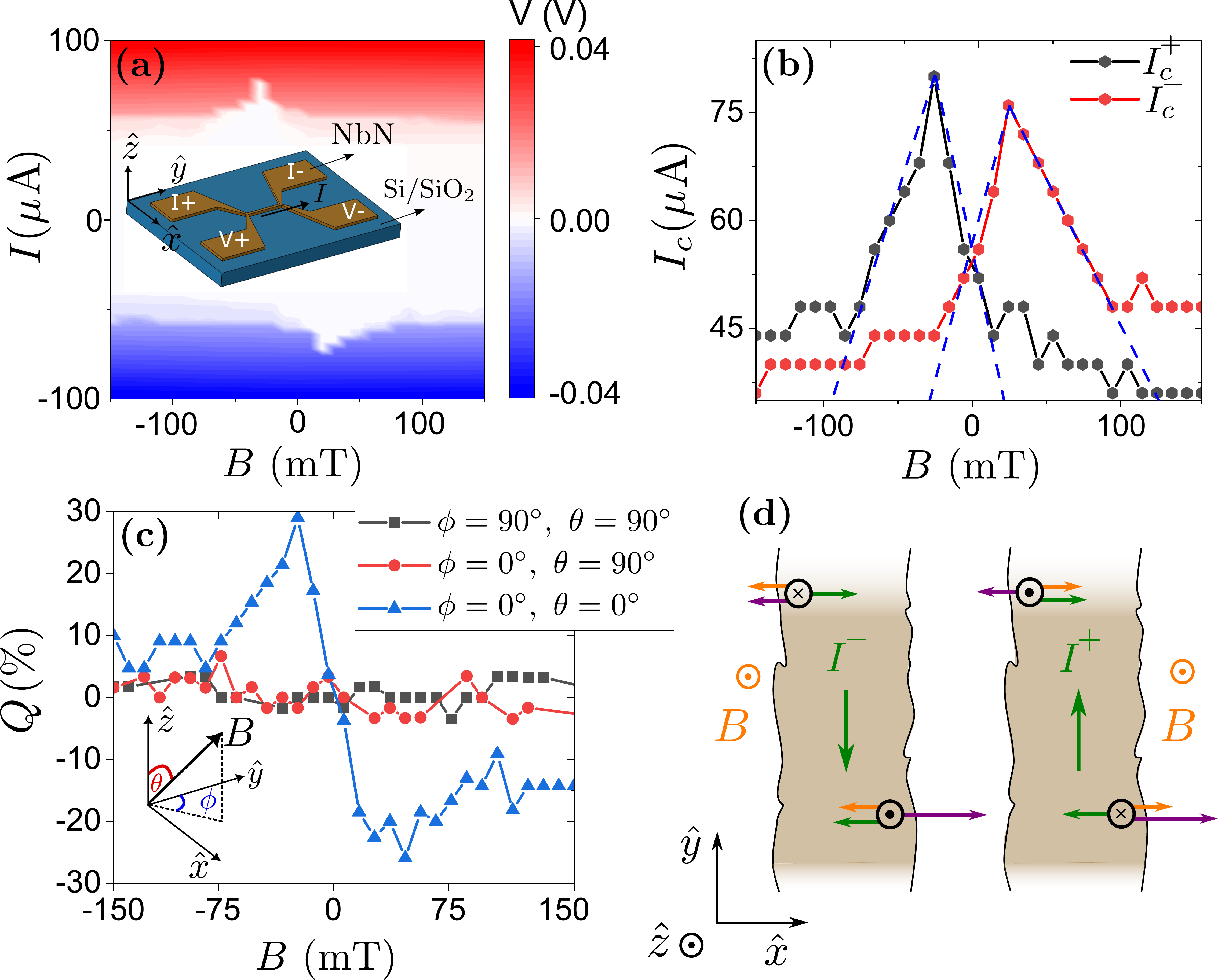}
\caption{(a) $IV$ curves of the device at T = 2 K, as a function of magnetic field in the out-of-plane $\hat{z}$ direction. Inset: schematic of the device used for experiments with the measurement configuration. (b) Critical current as a function of magnetic field extracted from fig.~a, for positive and negative bias  polarities. The precision of $I_c$ is limited by step-size of the current sweep; this is 5~$\mu$A in our measurements. (c)  Diode efficiency  versus magnetic field, for three configurations of $\theta$ and $\phi$ as specified in the legend.    $\theta~=~90^\circ$ and $\theta~=~0^\circ$ correspond to the configuration where the sample plane is parallel and perpendicular to the magnetic field, respectively. $\phi$ is the angle between $I$ and $B$, when the sample plane is parallel to the magnetic field.  $\phi~=~0^\circ$ and $\phi~=~90^\circ$ correspond to   $I~||~B$ and  $I~\perp~B$, respectively.  (d) Schematic depiction of the vortex instabilities at the two edges which determine the critical current. The arrows indicate forces  on the vortices due to the transport current (green), the Meissner current (orange) generated by the applied magnetic field (assumed positive), and the vortex surface barrier  (magenta). For transport current along $-\hat{y}$ (left panel), the critical current is determined by the left edge for $B \lesssim 20$ mT [corresponding to positive slope in (b)] and the right surface for $ 20~\mathrm{mT} \lesssim B \lesssim 100$ mT [negative slope in (b)]. For transport current along $\hat{y}$ (right panel), the  left edge determines the critical current at low $B$ ($> 0$). }   
\label{diode-01}
\end{figure}

Our experiments use SiO$_2$ (100 nm) on Si  and  epitaxial Y$_3$Fe$_5$O$_{12}$ (YIG) (100 nm) on gadolinium gallium garnet   as substrates. A positive photo-resist was spun on the substrates, which was then  patterned to micro devices via photo-lithography. The patterned sample was cleaned by soft sputtering Ar$^+$ ions in an ultra-high vacuum chamber, to eliminate any residual photo-resist and adsorbates. Further, NbN thin film was deposited   using  reactive DC magnetron sputtering at room temperature, at a base pressure $<$ 10$^{-8}$~mbar. The sample was then dipped in acetone to lift off the film, resulting in micro devices. A standard 4-point measurement configuration using a DC current source and a nano-voltmeter was employed for the measurements [inset of fig.~\ref{diode-01}~(a)].

We first present the results of experiments on the 20 nm thick NbN device on SiO$_2$ demonstrating non-reciprocity of the critical current, $I_c$. The device is a micro-bridge of lateral dimensions 1~$\mu$m $\times$ 4~$\mu$m.  The resistance versus temperature dependence of the device shows a broad transition to the superconducting state at $T_c$ = 7.5~K [refer to fig.~S3 in SI], which corresponds to a Bardeen-Cooper-Schrieffer energy gap 2$\Delta_0$ = 4.05~$k_{\text{B}}T_{\text c}$  equivalent to 1.3 meV, where $k_{\text B}$ is the Boltzmann constant.  The Ginzburg-Landau coherence length is estimated as \cite{DeSimoni2018} $\xi_{\text{GL}} = \sqrt{\hslash /\rho_{N} N_{\text F}e^2\Delta_0}~\approx~13.3 $~nm, where $N_{\text F} \approx   10^{28}/(m^{3}eV)$ is the density of states in NbN at the Fermi level  and $e$ is the electronic charge. This places the Pearl length~\cite{Pearl1964} ($  \lambda_{\text P} = 2 \lambda_{\text L}^2 / t$) at around 25 $\mu$m ($>> w$) ensuring a spatially uniform current through the film. 
Here $\lambda_{\text L}$ is the London penetration depth, derived using $\lambda_{\text L} = \sqrt{\hslash \rho_N /\pi \mu_0 \Delta_0}$, where $ l = 4~\mu$m, $w = 1~\mu$m, and $t = 20$~nm are length, width, and thickness of the bridge, respectively. $\rho_N$ is the resistivity in the normal (N) state at low temperature when resistance is 400~$\Omega$, and $\mu_0$ is the magnetic permeability in free space.

\begin{figure*}[tbh]
\centering
\includegraphics[width=16cm]{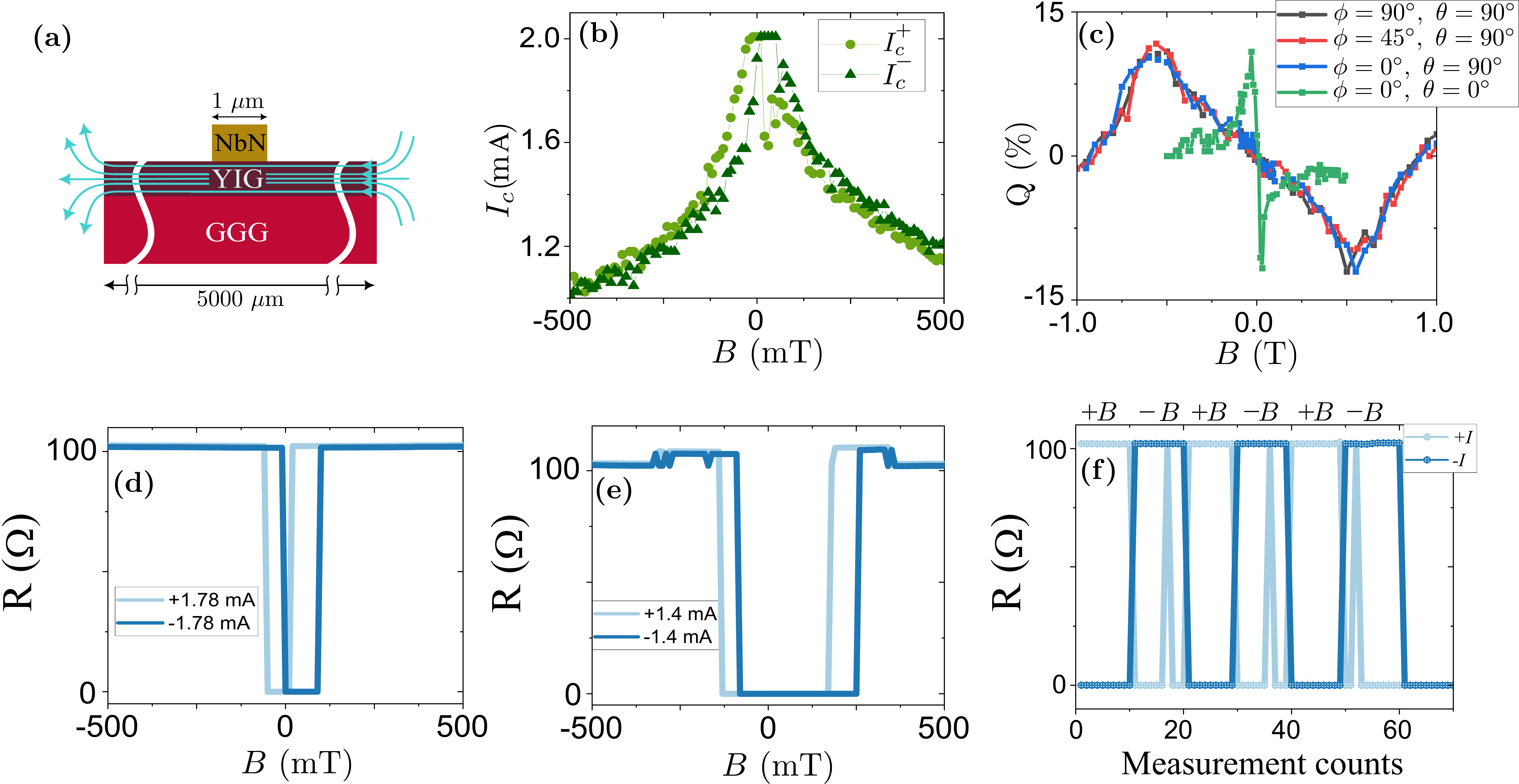}
\caption{ (a) Schematic  of the device cross-section. Lateral dimension of NbN ($\approx$~1~$\mu$m~$\times$~4~$\mu$m) is orders of magnitude smaller than YIG/GGG ($\approx$~5~mm $\times$ 5~mm); therefore the effect of fringe field on the micro-bridge is absent. The measurement geometry is the same as  in fig.~1.  YIG spreads over the whole substrate, making fringe field on the SC device negligible.  (b)  Critical current as a function of magnetic field for NbN/YIG device, for positive and negative bias polarities.   (c)   The diode efficiency   versus magnetic field plotted at different angles between direction of magnetic field and current as specified in the legend. (d), (e) Resistance versus magnetic field in the out-of-plane direction for currents in both bias polarities at (d) 1.78~mA  and  (e) 1.4~mA. (f) Resistance switching between SC  and normal  states  by changing magnetic field/bias polarity, showing robust diode effect at 7.3~K. The spikes in transition between SC and N states in the $+I$ curves are due to temperature fluctuations ($\approx$ 10~mK) which become visible close to $T_c$.}
\label{yig}
\end{figure*}

The current-voltage ($IV$) characteristics of the device as a function of magnetic field in the $\hat{z}$ direction exhibits a decay of $I_c$ with magnetic field up to $\approx$ 100~mT [fig.~\ref{diode-01}~(a)] and saturates for larger fields. We notice that the magnitude of maxima of critical currents ($\approx$ 80 $\mu$A) in the two bias polarities are asymmetric about $B = 0$,  indicating the emergence of non-reciprocity with respect to magnetic field. We plot the variation of $I_{c}^{+}$ and $I_{c-}^{-}$ as a function of magnetic field [fig.~\ref{diode-01}~(b)], and notice that $I_{c}^{+}$ $>$ $I_{c-}^{-}$, when $B~<~0$ and vice versa when $B~>~0$, demonstrating rectification. The figure of merit of the diode efficiency is defined by $Q =  \frac{I_{c}^{+} - I_{c-}^{-}}{I_{c}^{+} + I_{c}^{-}}$. The maximum efficiency of our device is $\approx$ 0.3 at T~=~ 2~K and  $B~\approx~\pm$~25~mT, illustrating a robust diode effect in the device.  To investigate the underlying mechanism causing the diode effect in NbN, we study the variation of $Q$ as a function of angle between the current and magnetic field directions. We find that the diode effect vanishes when  the applied magnetic field is in the plane of the sample, irrespective of the in-plane angle ($\phi$) [fig.~\ref{diode-01}~(c)]. The critical current as a function of out-of-plane magnetic field has been studied in NbN previously, however there was no report of non-reciprocity of the critical current \cite{ilin2014}. In our experiments, three out of seven devices did not show any SDE despite fabrication under identical conditions, implying that the feature is highly sample dependent. 

 We  interpret our findings assuming that the critical current is determined by the vortex flow~\cite{Shmidt1970,Shmidt1970b}. As discussed above, the critical current in our sample is the value at which the Lorentz force acting on out-of-plane vortices is able to overcome the net surface barrier and drive the vortices through the film. Following Ref.~\onlinecite{Hope2021}, we depict the forces on the relevant vortices near the left and the right edges in fig.~1~(d). Since the microstructure at the two edges  are never identical~\cite{Hope2021,vodolazov2005,sivakov2018}, we assume the surface barrier at the right edge to be larger. Let us first consider the case of transport current along $- \hat{y}$ [fig.~1~(d) left panel]. At zero $B$, the critical current is determined by the weaker left edge surface barrier. As $B~(>0)$ is increased, the Meissner screening currents caused by finite $B$ exert additional forces on the vortices, thereby reinforcing the left surface barrier [fig.~1~(d) left panel] and enhancing the critical current. Since the Meissner response reinforces the left surface barrier while weakening the right surface barrier, at $B \approx 20$ mT, the vortex instability determining the critical current shifts to the right edge. Thus, the critical current   decreases linearly with $B$ \cite{DeGennes1999,Shmidt1970,vodolazov2005,Hope2021}. On the other hand, for transport current along $+ \hat{y}$ [fig.~1~(d) right panel)], the left surface barrier is weakened by the Meissner currents leading to a linear decrease in the critical current. Combining these two cases, we see that the diode efficiency should increase linearly with $B$ at low fields, as observed in fig.~1~(c). At $B~\gtrsim$~50mT, the relevant vortex surface barrier has been lowered sufficiently such that bulk vortex pinning begins to determine the critical current. Therefore, the diode efficiency decreases, as the IS breaking caused by the surfaces starts to become irrelevant. The efficiency appears to saturate, instead of going to zero, possibly due to an internal and disorder-mediated IS breaking in the bulk pinning.

Complementing this qualitative analysis, we now extract the so-called maximum super-heating field of the Meissner state~\cite{Shmidt1970} $B_s$ from our recorded critical current dependence on $B$ [fig.~1~(b)]. This is the magnitude of $B$ at which the linear decrease in $I_c$ valid at low fields intercepts the magnetic field axis  [blue dashed lines in fig.~1~(b)], when the whole curve is shifted along the $B$ axis to make the critical current maximum occur at $B = 0$. We obtain 4 values for   $B_s$ (60 mT, 70 mT, 82 mT, 120 mT) from the four linear interpolations, each corresponding to the instability of a specific vortex (flux up or down) on a specific side (left or right). The difference in  values of $B_s$ probably arises due to  different local superconducting properties near the two edges. The extracted values of $B_s$ fit well with the theoretical order-of-magnitude estimate of $\sim 30$ mT obtained using the expression:~\cite{hou2022} $B_s = \phi_0 / (\sqrt{3}\pi\xi w)$, where $\phi_0$ is the flux quantum.~\footnote{The expression for $B_s$ has been evaluated as $B_s = \phi_0 / (2\pi\xi w)$ by several authors~\cite{Shmidt1970,Maksimova1998,Hope2021}. However, since it is only an order-of-magnitude estimate, this difference of $\sqrt{3}/2 \approx 0.9$ with respect to the expression employed in the main text is not important.} The agreement between the $B_s$ values obtained from our experiment and theory supports the validity of our assumed vortex mechanism of the critical current. Further confidence is gained from two observations: (i) variation of the critical current on a field scale of $B_s$, which is much smaller than the critical fields of NbN, and (ii) consistence between theoretical and experimental values of $B_s$ in our work as well as the experiments by Hou and coworkers, who recorded an order of magnitude smaller $B_s$ than ours due to  their much wider samples.

With the aim of examining the helical superconducting state   \cite{Edelshtein1989,Ando2020,daido2022,noah2022,shintaro2018} and the fringe field  mechanisms  of SDE, we study SC/magnetic insulator hybrids of NbN/YIG. Since our NbN film is grown on a much wider YIG film, our device design practically eliminates the effect of fringe fields on NbN  [fig.~2~(a)].  Diode effect in SC/ferromagnet hybrids has been reported previously, where the ferromagnet  is metallic \cite{carapella2019, papon2008,carapella2009} and insulating \cite{touitou2004}. We chose to study an SC/MI hybrid owing to its advantages for non-dissipative electronics, and to eliminate a parallel electron transport channel. In our NbN/YIG device, the critical current  exhibits non-reciprocity with diode efficiency factor $\approx$ 13\% [fig.~\ref{yig}~(b)]. In the configuration where the magnetic field is nominally in the plane of the sample, we observe that the  $Q (B)$ [fig.~\ref{yig}~(c)] is isotropic with respect to the direction between magnetic field and current. The  diode characteristics can be attributed to a possible misalignment of the sample with respect to the magnetic field, where the out-of-plane magnetic field is non-zero. Even a minor misalignment $\approx~3^{\circ}$ is sufficient to lead to the observed diode non-reciprocal behavior. Furthermore, isotropic behavior in the $\phi$ dependence corroborates this inference; any likely contribution from the Rashba spin-orbit coupling or magneto-chiral effects, would result in anisotropy in the in-plane field configuration \cite{davydova2022,bergeret2022,Baumgartner2022}. Thus, our experiments rule out any role of the spin-orbit coupling and associated helical superconducting states in causing the SDE in our samples.  Since our samples eliminate the fringe field effects, we observe basically zero SDE in our SC/MI sample under the in-plane field configuration. This supports  the fringe-field mechanism~\cite{hou2022} of SDE in samples with equal width of SC and MI layers (V/EuS), where the fringe fields affect the SC. The SDE in our SC/MI sample is due to the same vortex mechanism as discussed above, and we obtain  $B_s$ of $\sim~100$~mT from our experimental data [fig.~2~(b)]. Thus, the MI plays no important role in the SDE observed here.

Further, the switching between SC and N states is illuminated by analyzing the horizontal line-cut profiles of the surface plots [fig.~\ref{yig}~(d-e)] of $IV$ curve at different $B$ [refer to fig.~S1~(b) in SI]. The device exhibits $I_c~=~\pm 2$~mA at $B = 0$.  We  plot the resistance of the device versus $B$ at fixed values of current $|I|~<~|I_c|$  [fig.~\ref{yig}~(d-e)]. We  observe  switching between SC and N states, in addition to maintaining the  non-reciprocity with respect to  the two bias polarities.  The critical field $B_c$ (field at which the device transitions from SC to normal state), increases with a decrease in the current at which the resistance is measured.  Furthermore, we notice that $B_c$ is  asymmetric about $B = 0$. For instance, for a given bias polarity $+ 1.4$ mA [fig.~\ref{yig}~(e)], $B_c^{+} = 180$~mT and $B_c^{-} = -120$~mT. These observations are consistent with the correlation between $B_c (T)$ and $I_c (T)$ of a superconductor.  For each value of current, there exists a region of magnetic field where the device is superconducting in the positive bias   and normal in the opposite, and vice versa. Keeping  the magnetic field constant within  the regime of non-reciprocity, we can observe the canonical diode effect [fig.~\ref{yig}~(f)] over multiple measurement cycles. The resistance of the device  indicates switching from the SC state to the N state and vice versa by reversing the polarity of either the magnetic field or current bias. In the switching curve corresponding to $I+$ there are additional sporadic jumps to normal state, which primarily arise from  vortex instabilities due to small  temperature fluctuations. Over multiple measurement cycles we found that these jumps are random in nature, and absent at lower temperatures.  We observe magneto-resistance switching  up to 10~K [fig.~S2~(b) in SI], enhancing the temperature regime in which the SC diode based experiments can be performed. 

We have reported a robust superconducting diode in NbN and NbN/YIG micro devices, with diode efficiency  of $\approx$ 30\%.   Absence of rectification in the in-plane field with magnetically saturated YIG film provides complimentary evidence to the report of Hou et  al., \cite{hou2022} that fringe fields are responsible for their observations on V/EuS. 
 All our observations are consistent with the vortex surface barrier mechanism of the critical current. In our best devices we find the resistance switching persistent up to 10~K  which marks an advancement in the temperature range in which the diode based applications can be functional \cite{Lustikova2018}.

\vspace{0.15cm}

{\it Acknowledgments---} We thank J.~S.~Moodera, F.~S.~Bergeret and N.~Paradiso for insightful discussions.  AK acknowledges financial support from the Spanish Ministry for Science and Innovation -- AEI Grant CEX2018-000805-M (through the ``Maria de Maeztu'' Programme for Units of Excellence in R\&D). WB acknowledges funding by the Deutsche Forschungsgemeinschaft (DFG) through SFB 1432 (Project No. 425217212) and BE 3803/13 (Project No 465140728)
as well as from the European Union’s Horizon 2020 FET Open programme SuperGate (Grant Number 964398). MK, CB and CS acknowledge funding by the Deutsche Forschungsgemeinschaft (German Research Foundation), Project-ID 314695032—SFB 1277 (Subprojects: A08 and B08).

\newpage
\clearpage
\section*{Supplementary Information for the article\\Non-reciprocity of Vortex-limited Critical Current in Conventional Superconducting Micro-bridges}

\begin{figure}[tbh]
\centering
\includegraphics[width=\linewidth]{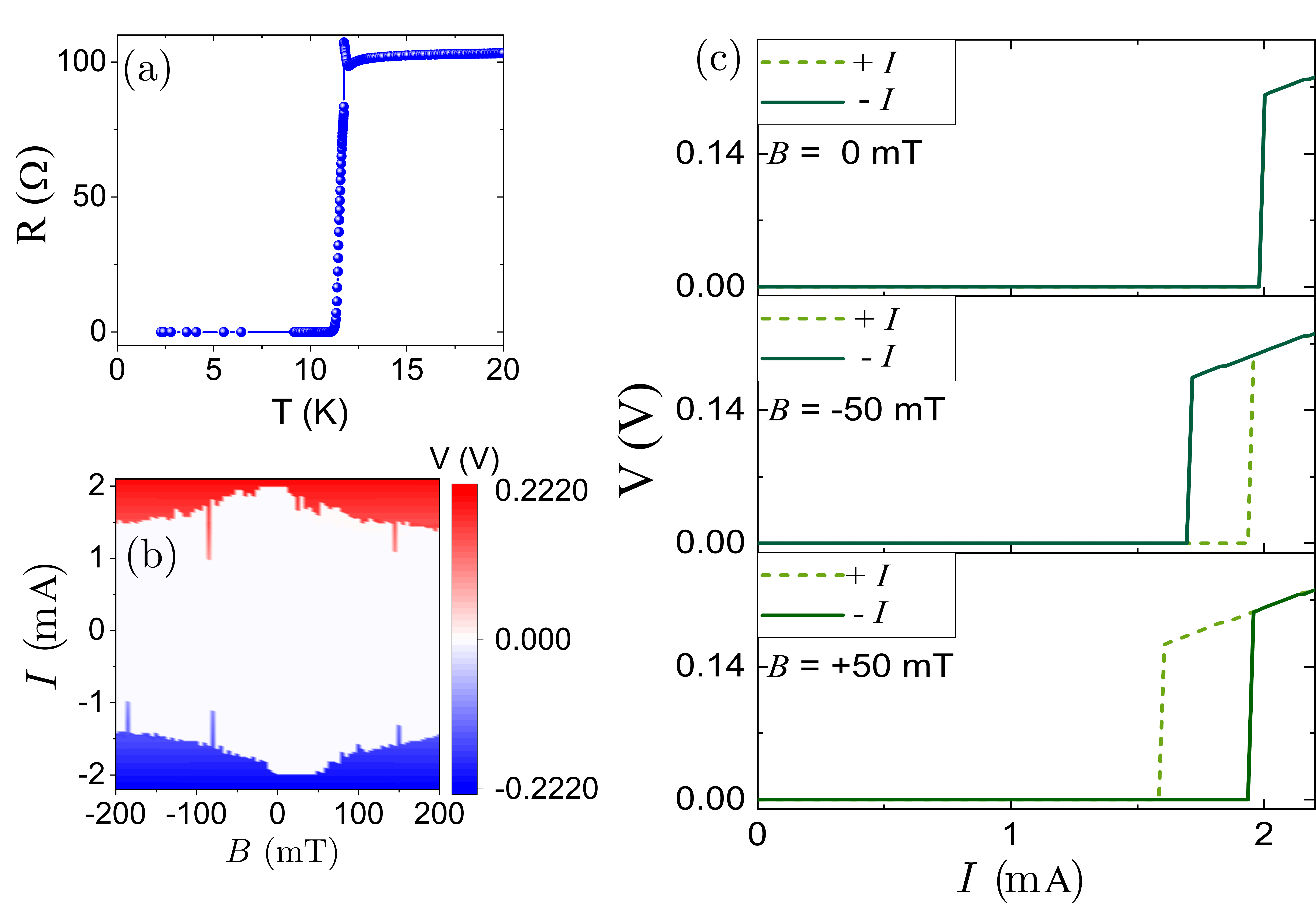}
\end{figure}

Fig. S1: (a) Resistance versus temperature of a 50 nm NbN/YIG device (b) $IV$ curve of the device at different magnetic fields  perpendicular to sample plane, at $T = 2$ K. (c) Absolute magnitude of the $IV$ curves in positive and negative bias directions at $B = 0,~B = -50$ mT and $B = + 50$ mT, showing non-reciprocal critical current.

Fig. S1 (a) shows the resistance versus temperature of the NbN/YIG device with $T_c$ $\approx$ 11~K. The pre-transition peak close to $T_{\text{c}}$ can be attributed to spatial inhomogeneities of the device as observed by earlier works on similar superconductors [Vaglio et al., Phys. Rev. B 47, 15302 (1993)]. In our device, we do not see any  influence of the pre-transition peak on the diode effect. The resistance of the normal state observed in the $IV$ curve is $\approx$~ 97~$\Omega$; where as the resistance of the peak is larger, implying that the device transitions into stable resistive state in the $IV$ curve.  Fig. S1 (b) shows the $IV$ curve of NbN/YIG device at different magnetic fields. We notice that the critical current is asymmetric with respect to magnetic field, $B = 0$. The sharp peaks in the $IV$ curve (at $\approx$ $-190$~mT, $-80$~mT and $+150$~mT) are outliers; we do not find them consistently reproduceable when the field sweep in repeated. Hence, they can be attributed to instabilities of the vortex distribution in the superconductor. To examine this carefully, we plot the IV curve at $B = 0$, $B = - 50$ mT and $B = + 50$ mT [fig. S1 (c)]. The critical current in the positive bias direction ($I_{c+}$) and negative bias direction ($I_{c-}$) are equal when $B = 0$. However, $I_{c+} > I_{c-}$ for $B < 0$ and vice versa for $B > 0$. This marks the SDE in NbN/YIG devices.

\begin{figure}[tbh]
\centering
\includegraphics[width=\linewidth]{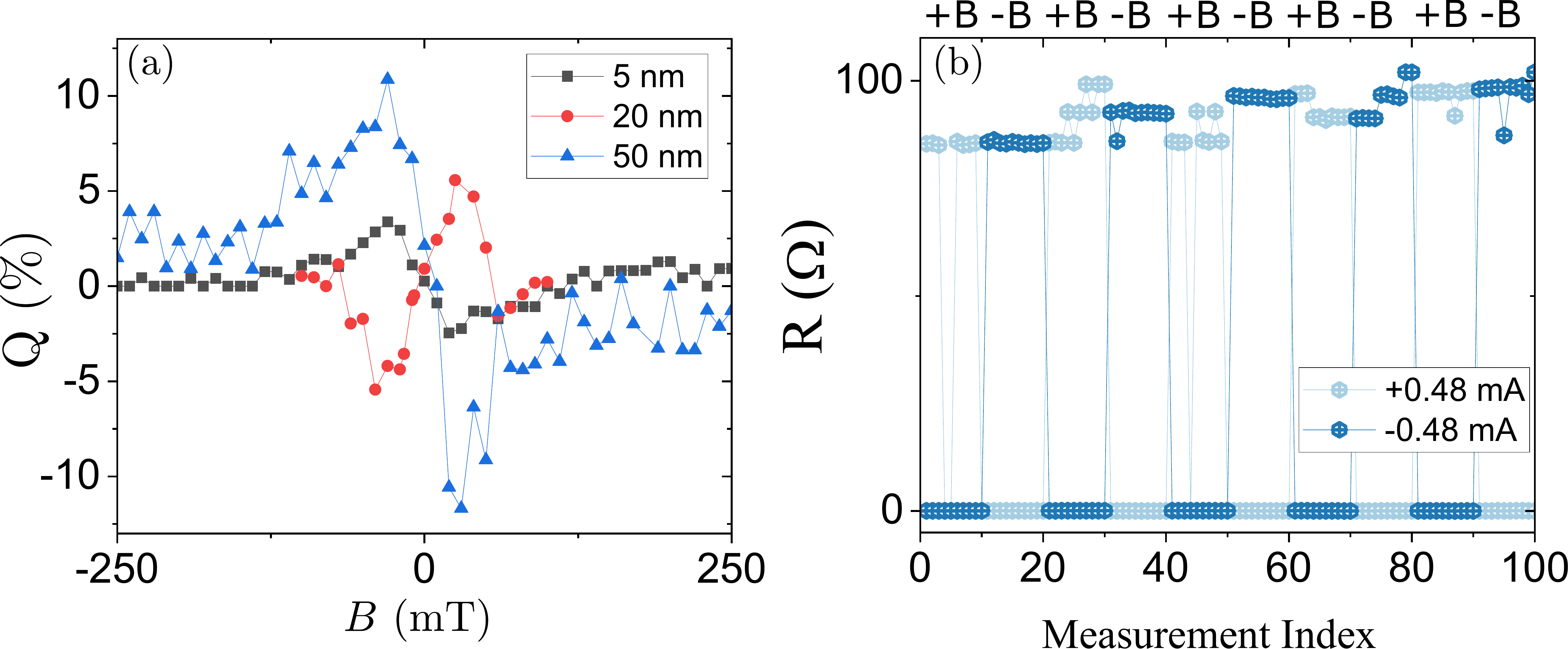}
\end{figure}
Fig. S2: (a) Diode efficiency factor for devices of varying thickness of NbN on YIG as shown in the legend, at $T = 2$ K. (b) Resistance of the device switching between superconducting state and normal state either by reversing bias or field polarity at $T = 10$ K.

Fig. S2 (a) shows $Q$ vs $B$ for devices of different thicknesses of NbN   on YIG, when the field is in the direction perpendicular to the sample. The nature of the diode effect is different due to non-identical sample quality in each device. From this data, the dependence of magnitude of $Q$ and the polarity on the thickness cannot be derived because the asymmetry in imperfections across the edges of the sample are not identical.    Fig. S2 (b) shows the resistance switching between normal state to the superconducting state by changing either the bias direction or the direction of magnetic field $B = + 50$ mT and $B = - 50$ mT, at $T = 10$~K. The ability to switch the superconducting device at $T = 10$~K enhances the temperature regime over which diode based applications can be performed.

\begin{figure}[tbh]
\centering
\includegraphics[width=8cm]{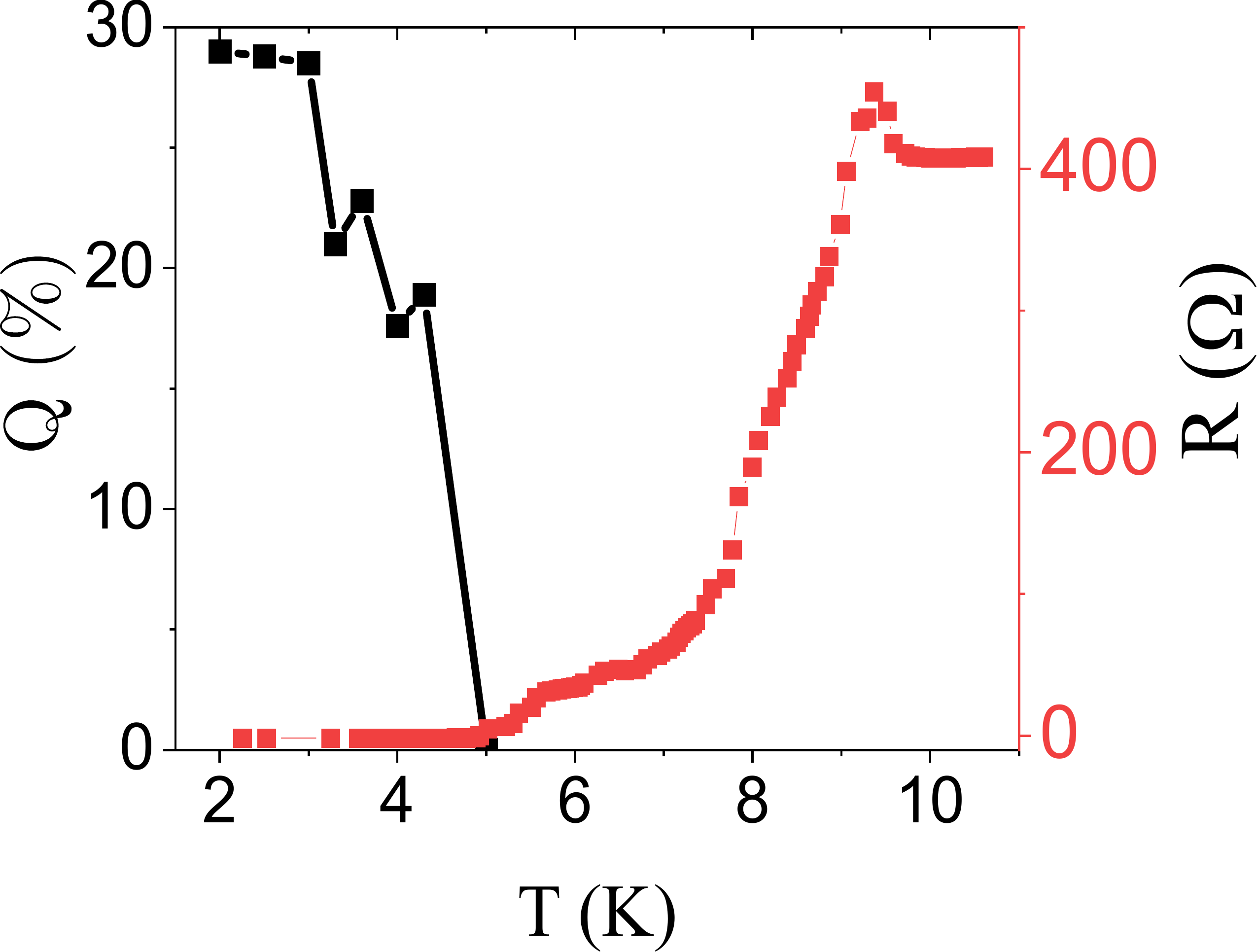}
\end{figure}
Fig. S3: Temperature dependence of the resistance of the device and the diode efficiency factor of a NbN/SiO$_2$ device.

Fig. S3 shows the temperature dependence of the resistance of the NbN device shown in the main text.  The critical temperature (7.3~K) is slightly smaller than that of the film on YIG due to lower thickness of the film, implying a higher degree of disorder that is also reflected  in a broader  transition curve to the SC state. The temperature dependence of the diode efficiency factor increases with decrease in temperature and saturates at low temperatures. A generalized model for temperature dependence  is of interest for further theoretical work.

\end{document}